# Domain Wall Motion Driven by Laplace Pressure in CoFeB-MgO Nanodots with Perpendicular Anisotropy

Yu Zhang,[1,2,3,a)] Xueying Zhang,[1,2,3,a)] Nicolas Vernier,[3] Zhizhong Zhang,[1,2] Guillaume Agnus,[3] Jean-Rene Coudevylle,[3] Xiaoyang Lin,[1,2] Yue Zhang,[1,2] You-Guang Zhang,[1,2] Weisheng Zhao,[1,2*] and Dafiné Ravelosona[3]

[1]*Fert Beijing Research Institute, BDBC, Beihang University, Beijing 100191, China*

[2]*School of Electronic and Information Engineering, Beihang University, Beijing 100191, China*

[3]*Centre de Nanosciences et de Nanotechnologies, University of Paris-Sud, CNRS, Orsay 91405, France*

**We have studied the magnetization reversal of CoFeB-MgO nanodots with perpendicular anisotropy for size ranging from w=400 nm to 1 μm. Contrary to previous experiments, the switching field distribution is shifted toward lower magnetic fields as the size of the elements is reduced with a mean switching field varying as 1/w. We show that this mechanism can be explained by the nucleation of a pinned magnetic domain wall (DW) at the edges of the nanodots where damages are introduced by the patterning process. As the surface tension (Laplace pressure) applied on the DW increases when reducing the size of the nanodots, we demonstrate that the depinning field to reverse the entire elements varies as 1/w. These results suggest that the presence of DWs has to be considered in the switching process of nanoscale elements and open a path toward scalable spintronic devices.**

## I. INTRODUCTION

Magnetic nanostructures based on perpendicular magnetic anisotropy (PMA) [1,2] materials are attracting a large amount of attention for their potential applications, such as high density magnetic random access memory (MRAM) [3,4,5], bit patterned media [6,7], magnetic logic [8,9,10], or in the fields of magnonics and spin waves [11,12]. The scalability of these applications toward ultimate technology nodes is in general limited by the structural variability of the nanostructures. This leads to a dispersion of the magnetic properties, which strongly affects the switching mechanism when the dimension of the nanostructures becomes smaller. In particular, this has been extensively shown for the switching process of magnetic dots [13,14,15,16,17,18]. When the size of the dots is sufficiently large, the dominant mechanism for switching has been found to be nucleation followed by rapid propagation of magnetic domain walls (DWs). In this case, as the propagation fields are usually lower than the nucleation fields [13,19], the Switching Field Distribution (SFD) corresponds to the distribution of nucleation fields, which is related to the distribution of magnetic anisotropy in the films. As the size of the dot decreases, the SFD is enlarged and shifted toward higher fields. A simple model taking into account the initial intrinsic distribution of magnetic anisotropy in the films can explained these results [14,15]. In addition to such variability of magnetic properties in the pristine films, edge damages introduced by the patterning process can also have a strong influence on the switching behavior [16]. This is the case for Spin-Transfer-Torque MRAM (STT-MRAM) or domain wall based nanodevices where the edges have been found to reduce the efficiency of the switching process at small dimensions [20,21,22].

---

a) Yu Zhang and Xueying Zhang contributed equally to this work.
* Author to whom correspondence should be addressed. Electronic mail: (W.Z.) weisheng.zhao@buaa.edu.cn

Here, we have used magneto-optic Kerr imaging microscopy [23,24] to study the magnetization reversal of magnetic nanodots based on CoFeB-MgO materials with PMA with sizes ranging from 400 nm to 1 μm. Contrary to previous results [15,16], we observe that the SFD is shifted towards lower magnetic fields when the size is reduced. Using the framework of elastic interface (e.g. soap bubbles), we show that the Laplace pressure applied on a DW nucleated and pinned at the edges of the nanodots is responsible for such mechanism.

## II. EXPERIMENTS METHODS AND RESULTS

CoFeB/MgO-based multilayers Ta (5 nm)/CuN (40 nm)/Ta (5 nm)/Co$_{40}$Fe$_{40}$B$_{20}$ (1.1 nm)/MgO (1 nm)/Ta (5 nm) were grown by sputtering using a Singulus TIMARIS cluster tool on 100-mm Si wafers. After the deposition, the samples were annealed in high vacuum at 380 °C for 20 min. Nanodots with different sizes were patterned by electron beam lithography (EBL) using a polymethylmethacrylate (PMMA) resist. Subsequently, a 50-nm-thick Al mask was deposited by electron beam evaporation onto the film and then lifted off and an ion milling process with Ar ions (etching angle of 45°) was used to pattern the magnetic dots. The magnetic properties of the continuous thin films were studied by using a Vibrating Sample Magnetometer (VSM) system. The switching process of the nanodots was investigated by using magneto-optic Kerr imaging microscopy with a wavelength of 450 nm [23]. All experiments were performed at room temperature.

**Fig. 1(a)** shows the hysteresis loop of the full films indicating a PMA. The coercivity of the CoFeB layer is as low as $\mu_0 H_c = 1$ mT, as we have shown in our previous study [19]. The effective anisotropy constant of our films is $K_{eff1} = 1.3 \times 10^5 \, J/m^3$ as calculated by VSM measurement using in-plane magnetic field. Considering the resolution limit of our magneto-optic Kerr microscope, the minimum size of nanodots fabricated in this work is 400 nm, as shown in **Fig. 1(b)**. The pitch (defined as the distance between the centers of two nanodots) was chosen to be 5 μm to minimize the influence of dipolar effects [25].

In order to characterize the switching process, the following procedure is employed: the film is first saturated with a large positive magnetic field along the perpendicular direction (easy axis) and then a negative magnetic field pulse with a duration of 1 ms is applied for investigating the switching process by Kerr microscope. The above procedure is repeated during the experiment and for each negative magnetic field pulse, the number of reversed islands is counted. The switching probability of the array is then obtained by calculating the ratio of the reversed nanodots to the overall number of the nanodots. **Figure 2** shows the typical magnetic switching process of the nanodots arrays for a size of 1 μm and 600 nm respectively. As expected, due to the variability of the nano-elements, a SFD is observed. For instance, for the 1 μm nanodots (see **Fig. 2(a)**), nearly 10% and 90% of the nanodots were reversed under magnetic fields of 10 mT and 16 mT, respectively. Surprisingly, we observe that for the array of 600 nm nanodots, the SFD is shifted to lower magnetic fields. In this later case, nearly 10% and 90% of the nanodots were reversed under magnetic fields of 9 mT and 12 mT respectively. The SFD measurements were repeated several times for each dimension and were reproducible within a variation of 2%.

The number of switched islands for nanodots size of 1 μm and 600 nm is presented in **Fig. 3(a) and 3(b)** respectively. In addition to the shift of the SFD to lower magnetic fields for smaller nanodots, we observe that the shape of the distribution is roughly unmodified. In order to fit the data, we have used the method based on integrated Gaussian distribution fits [16], where the cumulative distribution function (CDF) with an error function $erf(x)=\frac{2}{\sqrt{\pi}}\int_0^x e^{-t^2}dt$ is defined as:

$$P(H) = \frac{1}{2}\left[1+erf\left(\frac{H-H_{sf}}{\sigma\sqrt{2}}\right)\right] \qquad (1)$$

Then the average switching field (mean) $H_{sf}$ and the width of the distribution (standard deviation) $\sigma$ can be determined precisely, as shown in **Fig. 3(c)** and **Table I**. Consistent with **Fig. 2**, we observe a clear shift of the SFD toward lower fields when the size is reduced, without noticeable change for the width of the distribution $\sigma \sim 2$ mT. In addition, **Fig. 3(d)** indicates a linear relationship between the average switching field $H_{sf}$ and the inverse of the dot size w.

In order to understand why the SFD shifts to lower magnetic fields when reducing the size of the nanodots, the multilayers with identical stack are patterned into 400 μm large squares under the same fabrication process. As shown in **Fig. 4(a)**, after a magnetic field pulse of microseconds, although a few nucleation sites are present inside the squares as expected for much larger structures, we observe that most of the nucleation and propagation events occur along the edges. This result suggests the existence of a region of lower anisotropy at the edges of the elements that channels both DW nucleation and motion [29,30,31]. We believe that this feature is due to the patterning process [16,24,26,27,28,32,33], in particular the Ar ions milling that induces damages such as edge roughness, re-deposition on the sidewall, intermixing of the interfaces, or oxidation of the layers. Besides, owing to the difference in etching rate, material segregation of CoFeB may also lead to the different compositions of Co and Fe at the edges as well [19,23]. The damaged region is in general of the order of the grain size, which corresponds to a typical length scale of 10-20 nm [34,35].

## III. DISCUSSION

Below, we show that our results can be explained by the pinning of the DW at the edges of the nanodots together with the action of a Laplace pressure on the DW. First, as demonstrated previously [36], the presence of a gradient of anisotropy with a scale of the DW width Δ can induce strong DW pinning. In particular, if we consider a DW pinned at the edges of the nanodots, the depinning field is given by

$$\mu_0 H_{depin} = \frac{K_{eff1} - K_{eff2}}{2\cdot M_s} \cdot \frac{2\Delta}{\delta}\tanh\left(\frac{\delta}{2\Delta}\right) \qquad (2)$$

where $K_{eff1}$ and $K_{eff2}$ are the effective anisotropy in the non-damaged and damaged area respectively, δ the gradient length, $M_s$ the volume saturation magnetization and Δ the DW width. In particular, for δ with the same order of Δ, a variation of a few 10% of the anisotropy can give depinning fields of the order of a few 10 mT, which is much larger than the ultra-low intrinsic depinning fields of the films (~2 mT) [19]. As a result, once reversed domains are nucleated at the edges, they are expected to propagate only along the edges (i.e outside the pinning potential) as shown in **Fig. 4(a)**, and not toward the center of the nanodots (i.e cannot go across the pinning potential).

As we have evidenced recently, the Laplace pressure plays an important role in the dynamic of curved DWs [37]. This pressure originates from the DW surface tension [38,39,40], which is a mechanism quite well known for soap bubbles [41,42]. Once the curvature radius of a DW becomes smaller than several micrometers, this pressure can be high enough to dominate the dynamic of DWs, such as inducing the spontaneous collapse of a domain bubble. In the following, we further describe the depinning process to reverse the entire nanodot by introducing Laplace pressure on the DW. Due to the lower anisotropy induced in etching process, the DWs firstly nucleate at the edges of the dots. Note that the duration of the applied magnetic field pulse is 1 ms, which is sufficiently long for the nucleated DWs to propagate along the edges, even with a low speed in creep regime. Then the nucleated DWs are connected with each other and form a single DW along the edge of the dot. Note that in magnetic reversal, in addition to the driving magnetic field, a Laplace pressure is applied on the DWs with curvature (i.e an arc-shaped DW at the edge of the dots). This pressure favors the collapse of DW into the center of the dot and its strength is inversely proportional to the radius of curvature R of the DW. Considering a circular DW of radius R=w/2 at the edges of the nanodot, the Laplace pressure expresses as:

$$P_L = 2\lambda/w \tag{3}$$

where $\lambda$ is the interfacial energy density of the pinned DW. We note that for Bloch type domain walls, the interfacial energy density can be written $\lambda = 4\left(A_s \cdot K_{eff2}\right)^{1/2}$ where $A_s$ is the exchange stiffness constant. When an external magnetic field $H_{ext}$ is applied along the perpendicular direction, the Zeeman energy induces a pressure on the DW that can be written as:

$$P_{Hext} = 2 \cdot M_s \cdot \mu_0 H_{ext} \tag{4}$$

The Zeeman energy and the Laplace pressure act together to depin the domain wall as illustrated in **Fig. 4(b)**, which gives:

$$P_L + P_{Hext} = 2 \cdot M_s \cdot \mu_0 H_{depin} \tag{5}$$

By combining Equation (3), (4) and (5), this gives the minimum switching field $H_{sf}$ to overcome the pinning potential as:

$$\mu_0 H_{sf} = \mu_0 H_{depin} - \frac{\lambda}{M_s} \cdot \frac{1}{w} \tag{6}$$

This result is in agreement with the linear variation experimentally observed in **Fig. 3(d)** and indicates that the Laplace pressure just acts as a simple effective field proportional to 1/w, which only shifts the distribution without modifying its width $\sigma$. Using a linear fit allows us to determine ($\mu_0 H_{depin}$) and ($\lambda/M_s$). We find $\mu_0 H_{depin}$ = 16 mT and a DW energy $\lambda = 3.4 \times 10^{-3} J/m^2$, which is in very good agreement with previous studies [37,43]. Finally, by considering $\mu_0 H_{depin}$ =16 mT, a typical gradient of δ~Δ, the experimental values $K_{eff1} = 1.3 \times 10^5 J/m^3$ and $M_s = 1.3 \times 10^6 A/m$, Eq. (2) gives $K_{eff2} = 8.5 \times 10^4 J/m^3$. These results indicate that due to patterning induced damages, a gradient of anisotropy of about 30% is present at the edges of the nanodots on a length scale of the DW width.

## IV. CONCLUSION

In conclusion, we have demonstrated that due to the presence of edge damages, the DW nucleation and depinning process govern the field-induced magnetic reversal of nanodots. The Laplace pressure plays a critical role for explaining the decreased SFD with the reduced size of dots. Those features should be taken into account in advanced spintronic devices such as Spin-Orbit-Torque MRAM (SOT-MRAM) where the spin current can also induce the nucleation of DWs at the edges of the elements. These results also suggest a path toward scalable devices based on controlling the nucleation and pinning potential of DWs at the edges of the nano-elements. In this case, benefiting from the Laplace pressure and keeping the same thermal stability given by the gradient of anisotropy, a lower switching current would be needed when reducing the size of the devices.

The authors would like to thank Daoqian Zhu and Jiaqi Wei for technical support as well as constructive discussion. We gratefully acknowledge financial support from the European Union FP7 program (ITN WALL No. 608031), the French national research agency (COMAG, ELECSPIN) as well as International Collaboration Projects (No. 2015DFE12880 and No. B16001).

__________________________________________________

TABLE I. Average switching field and width of the distribution as a function of the dot size

| Dot Size nm | 1/Size μm$^{-1}$ | Average switching field [a] mT | Width of the distribution [a] mT |
|---|---|---|---|
| 400 | 2.50 | 9.5 | 2.2 |
| 600 | 1.67 | 11.5 | 2.3 |
| 800 | 1.25 | 12.6 | 2.2 |
| 1000 | 1.00 | 13.5 | 2.4 |

[a] *Average switching field $H_{sf}$* and the *width of the distribution $\sigma$* are obtained by extracting the parameters from experimental data (see Fig. 3).

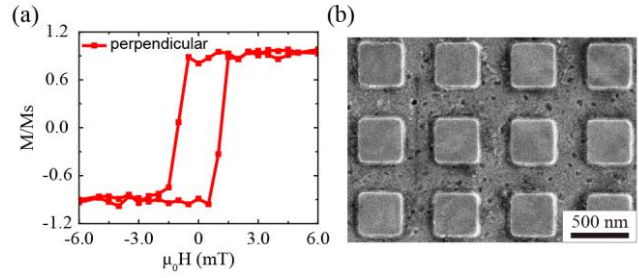

Fig. 1. (a) Hysteresis loops with out-of-plane magnetic field for Ta (5 nm)/CuN (40 nm)/Ta (5 nm)/Co$_{40}$Fe$_{40}$B$_{20}$ (1.1 nm)/MgO (1 nm)/Ta (5 nm) films. (b) SEM image of a patterned 400 nm nanodots array.

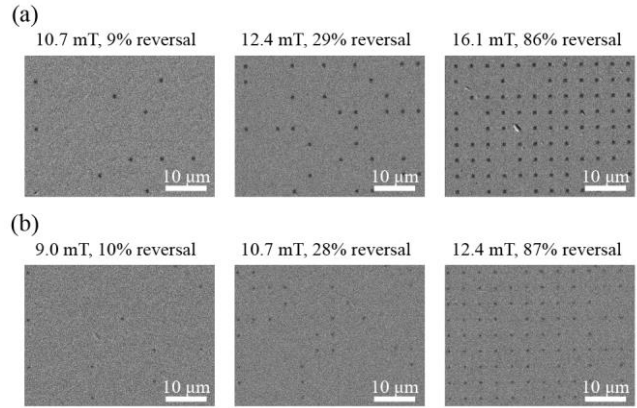

Fig. 2. Kerr microscopy images showing the switching process under magnetic fields for the nanodots array with sizes of (a) 1 μm and (b) 600 nm.

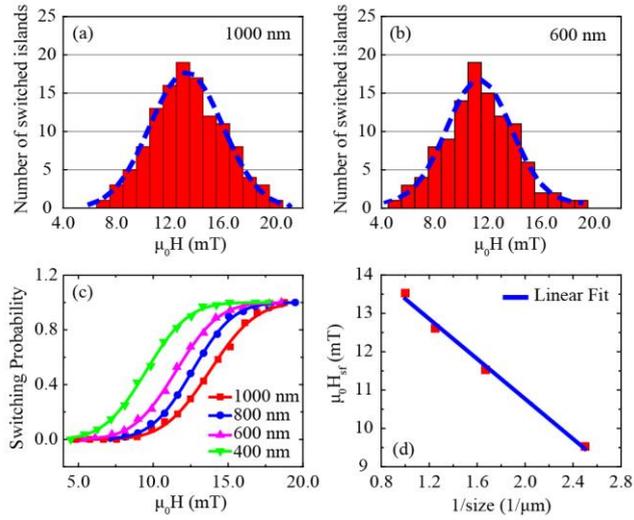

Fig. 3. Histogram indicating the number of islands that switch as a function of the applied magnetic field for a dot size of (a) 1 μm and (b) 600 nm. (c) Average switching probability as a function of magnetic field for dot sizes of 1 μm, 800 nm, 600 nm and 400 nm. (d) Average switching field as a function of the inverse of the dot size.

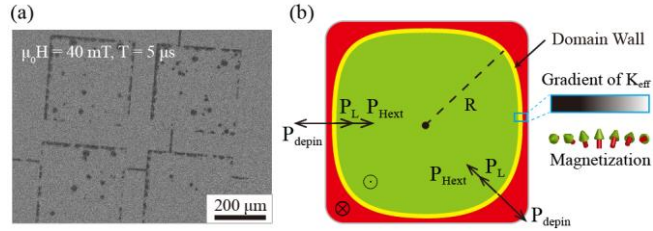

Fig. 4. (a) Kerr microscopy image of large squares (size of 400 μm) indicating the presence of nucleation events. The sample was first saturated with a strong magnetic field and then an opposite field of 40 mT was applied for 5 μs. (b) Schematic of the magnetization reversal process of a nanodot as described in the text. A single domain wall (yellow) is located at the edge of the dot, separating the reversed (red) and not reversed regions (green). The DW is pinned by a gradient of anisotropy on a length scale of δ~Δ due to edge damages. $P_{depin}$, $P_{Hext}$ and $P_L$ correspond to the pressures applied on the DW due to the pinning, the external magnetic field and the Laplace pressure, respectively.